\documentclass[12pt]{article}
\usepackage{epsfig}
\usepackage{amsfonts}
\usepackage{amsmath}
\usepackage[mathscr]{eucal}
\usepackage{cite}
\def\be{\begin{equation}}\def\ee{\end{equation}}
\def\Ga{\Gamma}  \def\eps{\epsilon}\def\lal{\mathcal{L}}\def\al{\alpha}

\def\bel{\begin{equation*}}\def\eel{\end{equation*}}
\def\w{\omega}

\begin{document}

\thispagestyle{empty}\rightline{\small DCPT-04/17 \hfill}
\rightline{\small hep-th/0405098 \hfill}
\vspace*{2cm}

\begin{center}
{\bf \LARGE Type IIB Killing spinors and calibrations}\\
\vspace*{1cm}

{\bf Emily~J.~Hackett-Jones}\footnote{E-mail: {\tt
e.j.hackett-jones@durham.ac.uk}} 
{\bf and Douglas~J.~Smith}\footnote{E-mail: {\tt
douglas.smith@durham.ac.uk}}

\vspace*{0.3cm} \vspace{0.5cm}
Centre for Particle Theory \\
Department of Mathematical Sciences \\
University of Durham, Durham DH1 3LE, U.K. \\

\vspace{2cm}
{\bf ABSTRACT}
\end{center}

In this paper we derive the full set of differential equations and some algebraic relations for $p$-forms constructed from type IIB Killing spinors. These equations are valid for the most general type IIB supersymmetric backgrounds which have a non-zero NS-NS 3-form field strength, $H$, and non-zero R-R field strengths, $G^{(1)}$, $G^{(3)}$ and $G^{(5)}$. Our motivation is to use these equations to obtain generalised calibrations for branes in supersymmetric backgrounds. In particular, we consider giant gravitons in $AdS_5\times S^5$. These non-static branes have an interesting construction via holomorphic surfaces in $\mathbb{C}^{1,2}\times\mathbb{C}^3$. We construct the $p$-forms corresponding to these branes and show that they satisfy the correct differential equations. Moreover, we interpret the equations as calibration conditions and derive the calibration bound. We find that giant gravitons minimise ``energy minus momentum''.

\vfill
\setcounter{page}{0}
\setcounter{footnote}{0}
\newpage

\section{Introduction}
Recently there has been much interest in classifying supersymmetric solutions of supergravity theories in various dimensions \cite{Figueroa-O'Farrill:2002ft,Gauntlett:2002fz,Gauntlett:2003wb,Cariglia:2004kk,Gauntlett:2004zh,Gauntlett:2003fk,Kaste:2003zd,Behrndt:2003uq,Gauntlett:2003cy,Dall'Agata:2003ir,Gowdigere:2003jf,Gutowski:2003rg,Caldarelli:2003pb,Pilch:2004yg,Dall'Agata:2004dk ,Frey:2004rn}. One technique which has proved particularly effective is to use the Killing spinors of the background to construct forms of different degrees. For example, the authors of Refs.~\cite{Gauntlett:2002fz,Gauntlett:2003wb} used $p$-forms, $\phi$, in the classification of general supersymmetric solutions of 11-dimensional supergravity. The components of $\phi$ are given by
\[
\phi_{M_1 \ldots M_p} = \bar\eps \Gamma_{M_1 \ldots M_p}\eps
\]
where $\eps$ is a Killing spinor of 11-dimensional supergravity and $\Gamma_M$ are Dirac matrices. These $p$-forms obey algebraic and differential relations descended from the Fierz identities and the Killing spinor equation, respectively. Moreover, the forms define a mathematical structure known as a G-structure, which is the reduction of the $Spin(10,1)$ frame bundle to a G-sub-bundle. The type of G-structure that arises 
can then be used to classify the supersymmetric solutions of 11-dimensional supergravity \cite{Gauntlett:2002fz,Gauntlett:2003wb}. Similar techniques have been used \cite{Gauntlett:2003fk,Cariglia:2004kk,Gauntlett:2003cy,Dall'Agata:2003ir,Gowdigere:2003jf,Gutowski:2003rg,Caldarelli:2003pb,Pilch:2004yg,Dall'Agata:2004dk ,Frey:2004rn} to (partially) classify supersymmetric solutions in various lower-dimensional supergravity theories.\\

As well as their use in classifying supersymmetric backgrounds, the forms constructed from Killing spinors are related to generalised calibrations for branes. For example, in Ref.~\cite{Martelli:2003ki} it was shown that generalised calibrations for M-branes naturally emerge from the differential equations satisfied by the forms. Here we will be interested in calibrations for branes in type IIB backgrounds. Some examples of generalised calibrations in particular type IIB backgrounds have been found \cite{Gauntlett:2001ur}. However, here we will be interested in finding calibrations for non-static probe branes, which has not been investigated previously. 
We begin by considering the most general supersymmetric backgrounds of type IIB supergravity. That is, we consider backgrounds which admit at least one Killing spinor and have background field strengths, $H, G^{(1)}, G^{(3)}$ and $G^{(5)}$ non-zero. 
We construct $p$-forms from the Killing spinors and derive the full set of differential equations for these forms. Some algebraic relations between the forms and the field strengths are also derived. These differential and algebraic equations could then be used in the classification of type IIB supersymmetric backgrounds, as demonstrated in Refs.~\cite{Pilch:2004yg,Dall'Agata:2004dk,Frey:2004rn} for some special classes of 10-dimensional backgrounds. However, our focus will be on using the forms, and their corresponding differential equations, to construct generalised calibrations for non-static D3-branes in IIB backgrounds. In particular, we will consider giant gravitons in $AdS_5\times S^5$.\\

Giant gravitons are non-static spherical branes in $AdS_5\times S^5$. The fact that they are non-static makes them an interesting example to consider from the point of view of calibrations, as most previous work on calibrations has involved static probe branes. An interesting construction of giant gravitons has been proposed by Mikhailov \cite{Mikhailov:2000ya}. In this construction the space $AdS_5\times S^5$ is embedded in ${\mathbb C}^{1,2}\times{\mathbb C}^3$. The giant graviton world-volume then arises from the intersection of a holomorphic surface in ${\mathbb C}^3$ with the embedded $S^5$. One of the benefits of constructing giant gravitons in this way is that the supersymmetry projection conditions become very simple. This is essentially because Killing spinors in $AdS_5\times S^5$ lift to covariantly constant spinors in ${\mathbb C}^{1,2}\times{\mathbb C}^3$, and consequently everything simplifies in the higher dimensional space.\\

The plan of this paper is as follows. In \S~\ref{pforms} we consider the gravitino Killing spinor equation for type IIB supergravity and we use it to derive differential equations for the forms. Then in \S~\ref{algebraic} we derive some algebraic identities for the forms using Fierz identities and the algebraic Killing spinor equation. In \S~\ref{complex}-\ref{susy} we discuss the Mikhailov construction of giant gravitons in some detail. Then in \S~\ref{diffgiants} the forms corresponding to these holomorphic giant gravitons are shown to obey the correct differential equations. In \S~\ref{calgiants} we consider the relationship between the differential equations derived in \S~\ref{pforms} and generalised calibrations.  
In particular, we are interested in probe D3-branes in backgrounds where the field strengths $H$, $G^{(1)}$ and $G^{(3)}$ are set to zero, which is the case for $AdS_5\times S^5$. We find a calibration bound for these branes and then show that the holomorphic giant gravitons saturate this bound in \S~\ref{holgiants}. These calibrated giant gravitons have minimal ``energy minus momentum'' in their homology class. Moreover, in \S~\ref{dualgiants} we show that dual giants also saturate the calibration bound and they minimise the same quantity as the ordinary giants. Our conclusions are given in \S~6.\\

\section{Differential equations for the $p$-forms}\label{pforms}
We begin by considering the Killing spinor equations for type IIB supergravity. Partial results \cite{Pilch:2004yg,Dall'Agata:2004dk,Frey:2004rn} have been obtained for backgrounds which preserve 4-dimensional Poincar\'{e} invariance. Also in Ref.~\cite{Gauntlett:2001ur} some differential conditions were derived as generalised calibrations for 5-branes wrapping special Lagrangian 3-cycles. However, the full set of equations for completely general type IIB backgrounds has not been given until now. Type IIB supergravity has two Killing spinor equations. One is algebraic, and arises from requiring that the variation of the axino-dilatino vanishes. The second equation is differential, and arises from varying the gravitino. In this section we will be interested in the second (gravitino) equation, and we will use it to compute derivatives of forms constructed from Killing spinors. In \S~\ref{algebraic} we will discuss algebraic relations between the forms, some of which can be obtained from the algebraic Killing spinor equation. We expect that both the differential and algebraic relations we derive will play an important role in the full classification of supersymmetric type IIB backgrounds.\\

Following Ref.~\cite{Papadopoulos:2003jk}, the gravitino Killing spinor equation in the string frame is $D_M \eps =0$, where $\eps$ is a 32-dimensional chiral spinor, with two 16-dimensional components\footnote{Strictly speaking, $\eps^1$ and $\eps^2$ are 32-component spinors with positive chirality with respect to $\Gamma_{11} = \Gamma^{0\ldots 9}$. However, the condition $\Gamma_{11} \eps^i = \eps^i$ reduces the number of non-zero components of each $\eps^i$ to 16, and so we consider $\eps^i$ to be 16-dimensional.}, i.e.
\[
\eps = \left(\begin{array}{c}
\eps^1\\
\eps^2\end{array}\right)
\] 
and\footnote{Our conventions are that the Dirac matrices satisfy $\{ \Gamma^A, \Gamma^B\} = 2 \eta^{AB}$, the metric has the signature $(-, +, +, \ldots )$ and $\eps_{0\ldots 9}=+1$. Note that since we only consider even products of $\Gamma$ matrices, we can consider the $\Gamma$ matrices to be $16\times 16$, rather than $32\times 32$ (the spin representation of the $\Gamma$ matrices decomposes into 16-dimensional blocks).}
\be
D_M = \nabla_M + \frac{1}{8} H_{M A_1 A_2} \Ga^{A_1 A_2}\otimes \sigma_3 + \frac{1}{16}e^{\phi} \sum_{n=1}^5 \frac{(-1)^{n-1}}{(2n-1)!} G_{A_1 \ldots A_{2n-1}}\Gamma^{A_1 \ldots A_{2n-1}}\Ga_M \otimes \lambda_n \label{deriv}
\ee
Here $\phi$ is the dilaton and $\nabla$ is the Levi-Civita connection. The matrices $\lambda_n$ are defined as follows
\begin{equation}
\lambda_n = 
\begin{cases}
\sigma_1 & \text{if $n$ even,}\\
i\sigma_2 & \text{if $n$ odd.} \label{lambdan}
\end{cases}
\end{equation}
where $\sigma_i$, $i=1,2,3$, are the usual Pauli matrices. The field strength $H$ in Eq.~(\ref{deriv}) is the NS-NS 3-form, related to a 2-form gauge potential, $B$, by $H=dB$.
The field strengths $G$ are defined by
\[
G^{(2n +1)} = dC^{(2n)} - H\wedge C^{(2n-2)}
\]
where $C^{(n)}$ are the Ramond-Ramond potentials.
These field strengths are not all independent, but $G^{(7)} = - *G^{(3)}$, $G^{(9)}= *G^{(1)}$ and $G^{(5)}$ is self-dual ($G^{(5)} = *G^{(5)}$).\\

We now construct $p$-forms of different dimensions from the 16-dimensional component spinors, $\eps^1$ and $\eps^2$, of a single Killing spinor $\eps$.
The components of a generic $p$-form, $\omega^{ij}$, are given by
\be
\omega^{ij}_{M_1 \ldots M_p} = \bar\eps^i \Gamma_{M_1 \ldots M_p} \eps^j \label{generalp}
\ee
where $i,j = 1,2$ and $\bar\eps^i= (\eps^i)^t \Gamma^0$. In general, this construction will produce $2\times 2$ matrices of forms for each dimension $p$. However, due to the chirality of the spinors, many $p$-forms are automatically zero. The components of the possible non-zero forms are 
\begin{itemize}
\item 1-forms 
\[
K^{ij}_M = \bar\eps^i \Ga_M \eps^j
\]
\item 3-forms
\[
\Phi^{ij}_{MNP} = \bar\eps^i \Ga_{MNP} \eps^j\qquad \text{for $i\neq j$,}
\]
\item 5-forms
\[
\Sigma^{ij}_{MNPQR} = \bar\eps^i \Ga_{MNPQR} \eps^j
\]
\end{itemize}
It is also possible to construct some higher-dimensional forms (two 7-forms, $\Pi^{ij}$, for $i\neq j$, and four 9-forms, $\Omega^{kl}$). However, these are simply dual to the lower-dimensional forms; $\Pi^{ij} = - *\Phi^{ij}$, $\Omega^{kl} = *K^{kl}$. Note also that the 5-forms, $\Sigma^{ij}$, are self-dual. Moreover, there exist relations between the ``off-diagonal'' forms, namely
\be
K^{12} = K^{21}, \qquad \Phi^{12} = - \Phi^{21}, \qquad \Sigma^{12} = \Sigma^{21}
\ee
These relations can be easily proved by computing the transpose of the components of each form. This means that there are only 7 ``independent'' forms to consider: $K^{11}$, $K^{22}$, $K^{12}$, $\Phi^{12}$, $\Sigma^{11}$, $\Sigma^{22}$ and $\Sigma^{12}$. Actually these forms are not independent, since they obey complicated algebraic relations (some of which we will derive in the next section).

We now compute the covariant derivatives of these forms. For each $p$-form, $\omega^{ij}$, whose components are given in Eq.~(\ref{generalp}), we will compute
\begin{eqnarray}
\nabla_N \omega^{ij}_{M_1 \ldots M_p} &=& \nabla_N (\bar\eps^i \Gamma_{M_1 \ldots M_p} \eps^j ) \label{covderiv}\\
&=& (\overline{\nabla_N \eps^i}) \Gamma_{M_1 \ldots M_p} \eps^j + \bar\eps^i \Gamma_{M_1 \ldots M_p} (\nabla_N \eps^j) \nonumber
\end{eqnarray}
The idea is to use the Killing spinor equation, $D_M \eps =0$, to replace $\overline{\nabla_N \eps^i}$ and $\nabla_M \eps^j$ with terms involving the fields strengths, metric and dilaton. The second step will be to antisymmetrize over the indices $N,M_1,\ldots M_p$ to obtain the ordinary derivative of $\w^{ij}$, i.e. $d\w^{ij}$. Our motivation for doing this is to obtain calibration conditions for branes in type IIB supersymmetric backgrounds.\\

The computations for the covariant derivatives of the forms are messy, so we do not present all the details here. However, a useful result which helps to simplify the expressions is the following: given a $q$-form, $v$, and $p$-form, $w$, where $q>p$,
\[
\iota_{*v} * w = (-1)^{p(q-p) +1}\iota_w v 
\]
Here $(\iota_w v)_{N_1 \ldots N_{q-p}} \equiv \frac{1}{p!} w^{M_1 \ldots M_p}\ v_{M_1 \ldots M_p N_1 \ldots N_{q-p}}$.
This is also consistent with the relation
\[
** w = (-1)^{p(10-p) +1} w
\]
We now present the results for the ordinary derivatives of the forms. While the equations look complicated, they are valid for the most general supersymmetric backgrounds which have non-zero field strengths, $H$, $G^{(1)}$, $G^{(3)}$ and $G^{(5)}$. 
Starting with the 1-forms, $K^{ij}$, we have
\be
dK^{11} = - \iota_{K^{11}} H + \frac{e^\phi}{2}\Big\{-  \iota_{G^{(1)}}\Phi^{12} + \iota_{G^{(3)}} \Sigma^{12}
+ \iota_{K^{12}} G^{(3)} -  \iota_{\Phi^{12}} G^{(5)}\mbox{\hspace{0.5em}}\Big\} \label{K11}
\ee
The equation for $K^{22}$ can be obtained from Eq.~(\ref{K11}) by replacing 
\[dK^{11}\rightarrow dK^{22}, \qquad \iota_{K^{11}} H \rightarrow -\iota_{K^{22}} H,\]with all other terms remaining the same.
For $K^{12}$ we obtain,
\be
(d K^{12})_{MN} = \frac{1}{2} {H^{A_1 A_2}}^{}_{[M}\Phi^{12}_{N]A_1 A_2} + \frac{e^\phi}{4}\Big\{ \iota_{G^{(3)}}(\Sigma^{11} + \Sigma^{22}) + \iota_{(K^{11} + K^{22})}G^{(3)}\Big\}_{MN} \label{K12}
\ee
The equation for $K^{21}$ is exactly the same as above since $K^{21} = K^{12}$.  For the 1-forms it is also interesting to calculate $\nabla^{}_{(M}K^{ij}_{N)}$, i.e. symmetrizing over the indices. If $\ \nabla^{}_{(M}K^{ij}_{N)} = 0\ $ then $K^{ij}$ corresponds to a Killing vector. In fact, we find that only the combination $K^{11} + K^{22}$ is Killing, i.e.
\[
\nabla_{(M}(K^{11} + K^{22})_{N)} = 0 
\] 
As we will see in \S~\ref{calgiants}, the combination $K^{11} + K^{22}$ appears naturally in the calibration bound for D3-branes. This is perhaps not surprising as the above equation shows that it corresponds to a symmetry of the metric.\\

The 3-form $\Phi^{ij}$ is non-zero only when $i\neq j$. The differential equation satisfied by $\Phi^{12}$ is
\begin{eqnarray}
(d\Phi^{12})_{MNPQ} =&& {H^{A_1 A_2}}^{}_{[M}\Sigma^{12}_{NPQ] A_1 A_2} + \frac{3}{2} K^{12}\wedge H \nonumber\\
&+&\frac{e^{\phi}}{2}\Big\{\iota_{G^{(1)}} (\Sigma^{11} + \Sigma^{22}) - \iota_{(K^{11}+K^{22})}G^{(5)} + \frac{1}{2} (K^{11} - K^{22})\wedge G^{(3)} \nonumber\\
&&\qquad\qquad + G^{(3)}_{A_1 A_2[M}(\Sigma^{22} - {\Sigma^{11})_{NPQ]}}^{A_1 A_2}\Big\} \label{phi12}
\end{eqnarray}
where the omitted indices are understood to be $\scriptstyle [MNPQ]$. Since $\Phi^{21} = - \Phi^{12}$ we do not need to work out the equation for $\Phi^{21}$ separately. 
For the 5-form $\Sigma^{11}$ we obtain the following differential equation
\begin{eqnarray}
(d\Sigma^{11})_{MNPQRS} &=& - \frac{15}{2} {H^A}^{}_{[MN}\Sigma^{11}_{PQRS]A} \nonumber\\
&& + \frac{e^{\phi}}{2} \Big\{ 2  K^{12} \wedge G^{(5)} + 2 G^{(1)}\wedge \Sigma^{12} -  3 \iota_{G^{(1)}} \Pi^{12}\nonumber\\
&&\qquad +\ 3 \iota_{G^{(3)}} \Omega^{12} - 15 G^{(3)}_{A[MN}{{\Sigma^{12}}_{PQRS]}}^A \nonumber\\
&& \qquad   +\ 15 {\Phi^{12}}_{A[MN} {G^{(5)}_{PQRS]}}^A - 12 G^{(3)}_{A_1 A_2 [M}{{\Pi^{12}}_{NPQRS]}}^{A_1 A_2}\Big\} \label{Sigma11}
\end{eqnarray}
where in Eqs.~(\ref{Sigma11})-(\ref{Sigma21}) the omitted indices are understood to be ${\scriptstyle [MNPQRS]}$. The differential equation for $\Sigma^{22}$ is\begin{eqnarray}
(d\Sigma^{22})_{MNPQRS} &=& \frac{15}{2} {H^A}^{}_{[MN}\Sigma^{22}_{PQRS]A} \nonumber\\
&& + \frac{e^{\phi}}{2} \Big\{ - 2  K^{12} \wedge G^{(5)} - 2 G^{(1)}\wedge \Sigma^{12} - 3 \iota_{G^{(1)}} \Pi^{12} \nonumber\\
&& \qquad +\ 3 \iota_{G^{(3)}} \Omega^{12} - 15 G^{(3)}_{A[MN}{{ \Sigma^{12}}_{PQRS]}}^A \nonumber \\ 
&& \qquad +\ 15{\Phi^{12}}_{A[MN} {G^{(5)}_{PQRS]}}^A + 12 G^{(3)}_{A_1 A_2 [M}{{\Pi^{12}}_{NPQRS]}}^{A_1 A_2}\Big\} 
\end{eqnarray}
and the equation for $\Sigma^{12}$ is 
\begin{eqnarray}
(d\Sigma^{12})_{MNPQRS} &=& \frac{3}{2} {H^{A_1 A_2}}_{[M}{\Pi^{12}}_{NPQRS]A_1 A_2} - \frac{3}{2} H\wedge \Phi^{12} \nonumber\\
&&+\ \frac{e^{\phi}}{4} \Big\{ 2 (K^{22} - K^{11})\wedge G^{(5)} + 2 G^{(1)} \wedge (\Sigma^{22} - \Sigma^{11})\nonumber\\
&&\qquad +\ 3 \iota_{G^{(3)}} (\Omega^{11} +\Omega^{22})- 15 G^{(3)}_{A[MN}{(\Sigma^{22} + \Sigma^{11})^{}_{PQRS]}}^A\Big\}\label{Sigma21}
\end{eqnarray}
Since $\Sigma^{21} = \Sigma^{12}$ the differential equation for $\Sigma^{21}$ is the same as above.\\


The 7-forms $\Pi^{ij}$ are non-zero only when $i\neq j$. The differential equation for $\Pi^{12}$ is given by
\begin{eqnarray}
(d \Pi^{12})_{M\dots U}& = &2 H_{A_1 A_2 [M}{\Omega^{12}_{N\dots U]}}^{A_1 A_2} + 28 H_{A[MN}{\Pi^{12}_{P\dots U]}}^A- \frac{3}{2} H \wedge \Sigma^{12}\nonumber\\
&& + \frac{e^{\phi}}{12} \Big{\{} 36 G^{(3)}_{A_1 A_2[M} {(\Omega^{22} - \Omega^{11})_{N\dots U]}}^{A_1 A_2} + 3 G^{(3)} \wedge( \Sigma^{11} - \Sigma^{22})\nonumber\\
&&\qquad + 12 \iota_{G^{(1)}} (\Omega^{11} + \Omega^{22}) - 14 G^{(5)}_{A_1 A_2 A_3 [MN} {(\Omega^{11} + \Omega^{22})_{P\dots U]}}^{A_1 A_2}\Big\}
\end{eqnarray}
where the omitted indices are understood to be ${\scriptstyle [MNPQRSTU]}$. Since $\Pi^{21} = - \Pi^{12}$ we do not need to work out the equation for $\Pi^{21}$ separately.
Finally, for the 9-forms $\Omega^{ij}$ we have
\begin{eqnarray}
(d\Omega^{11})_{M\dots W} &=& - 45 H_{A[MN} {{\Omega^{11}}_{P\dots W]}}^A\nonumber\\
&& \frac{e^{\phi}}{2} \Big\{ 4 G^{(1)} \wedge \Omega^{12} - 2 G^{(3)} \wedge \Pi^{21} - 135 G^{(3)}_{A[MN} {\Omega^{12}_{P\dots W]}}^A\nonumber\\
&&\qquad + 105 G^{(5)}_{A[MNPQ}{\Pi^{21}_{R\dots W]}}^A - 60 G^{(5)}_{A_1 A_2[MNP}{\Omega^{21}_{Q\dots W]}}^{A_1 A_2}\Big\} 
\end{eqnarray}
where the omitted indices here and in Eqs.~(\ref{Omega22})-(\ref{Omega12}) are understood to be ${\scriptstyle[MNPQRSTUVW]}$. For $\Omega^{22}$ we have 
\begin{eqnarray}
(d\Omega^{22})_{M\dots W} &=& 45 H_{A [MN}{{\Omega^{22}}_{P\dots W]}}^A\nonumber \\
&& \frac{e^{\phi}}{2} \Big\{ - 4e^{\phi} G^{(1)}\wedge \Omega^{12} - 2 G^{(3)}\wedge \Pi^{12} - 135 G^{(3)}_{A[MN}{{\Omega^{12}}_{P\dots W]}}^A\nonumber\\
&&\qquad - 105 G^{(5)}_{A[MNPQ}{{\Pi^{12}}_{R\dots W]}}^A + 6 G^{(5)}_{A_1 A_2[MNP}{{\Omega^{12}}_{Q\dots W]}}^{A_1 A_2}\Big\}\label{Omega22}
\end{eqnarray}
and finally
\begin{eqnarray}
(d\Omega^{12})_{M\dots W} &=& - \frac{3}{2} H\wedge \Pi^{12}\nonumber\\
&& + \frac{e^{\phi}}{4} \Big\{ 4 G^{(1)} \wedge (\Omega^{22} - \Omega^{11}) - 135 G^{(3)}_{A[MN} {(\Omega^{22} + \Omega^{11})_{P\dots W]}}^A \nonumber\\
&& \qquad\qquad  - 60 G^{(5)}_{A_1 A_2 [MNP} {(\Omega^{22} - \Omega^{11})_{Q\dots W]}}^{A_1 A_2}\Big\} \label{Omega12}
\end{eqnarray}
Since $\Omega^{21} = \Omega^{12}$ the equation for $\Omega^{21}$ is exactly as above. Eqs.~(\ref{K11})-(\ref{Omega12}) give the full set of differential equations obtainable from the gravitino Killing spinor equation for a general supersymmetric type IIB background.
\newpage
\section{Algebraic Relations}\label{algebraic}

There are two ways to obtain algebraic relations between the forms. The first way is to use Fierz identities. There are many possible Fierz identities for the Dirac matrices in 10-dimensions. However, here we will be interested in one particular class of identities given by~\cite{Kennedy:kp}
\be
\left(\Gamma^{(l)\ A_1 \ldots A_l}\right)_{\al\beta} \left(\Gamma^{(l)}_{A_1\ldots A_l}\right)_{\gamma\delta} 
= \sum_{k=0}^{10} a_{lk} \left(\Gamma^{(k)\ B_1 \ldots B_k}\right)_{\al\delta} \left(\Gamma^{(k)}_{B_1\ldots B_k}\right)_{\gamma\beta} \label{Fierz}
\ee
where $\al, \beta, \gamma,\delta$ are spinor indices and the coefficients $a_{lk}$ are given explicitly by
\[
a_{lk} = \frac{l!}{16\cdot k!} (-1)^{\frac{ (l+k)^2 - l - k}{2}}\
\sum_{p={\rm max}\{ 0, l+k-10\}}^{{\rm min}\{k,l\}} (-1)^p \left( \begin{array}{c}
10-k\\
l-p\end{array}\right)\left(\begin{array}{c}
k\\
p\end{array}\right)
\]
These identities will allow us to find relationships between $K^{ij}\cdot K^{kl}$, $\Phi^{ij}\cdot\Phi^{kl}$ and $\Sigma^{ij}\cdot\Sigma^{kl}$, where $i,j,k,l\in \{1,2\}$ and we define
\begin{eqnarray*}
\Phi^{ij}\cdot \Phi^{kl} &=& \frac{1}{3!} \Phi^{ij}_{A_1 A_2 A_3}\ \Phi^{kl\ A_1 A_2 A_3}\\
\Sigma^{ij}\cdot \Sigma^{kl} &=& \frac{1}{5!} \Sigma^{ij}_{A_1 \ldots A_5}\ \Sigma^{kl\ A_1 \ldots A_5}
\end{eqnarray*}
In fact, somewhat surprisingly, these Fierz identities give 
\be
K^{ij} \cdot K^{kl} = 0, \qquad \Phi^{ij}\cdot \Phi^{kl} = 0, \qquad \Sigma^{ij}\cdot \Sigma^{kl} = 0
\ee
This is different to the 11-dimensional case. In 11 dimensions the Killing vector $K$ (which corresponds to $K^{11} + K^{22}$ here) can be time-like or null \cite{Gauntlett:2002fz,Bryant}. Here $K$ can only be null. Moreover, since each $K^{ij}$ is null and all scalar products vanish, this means that all $K^{ij}$ are proportional to the same null vector, i.e. $K^{ij} = c^{ij} \tilde{K}$, for some constants $c^{ij}$. Note that $\Phi^{ij}$ and $\Sigma^{ij}$ are also null and all scalar products between forms of the same degree vanish. We find the same results using the $\Gamma$-matrix algebra package GAMMA \cite{Gran:2001yh}. Presumably there are other non-trivial algebraic relations which could be obtained by considering other types of Fierz identities (e.g. $K\wedge \Phi$ and $\iota_K \Sigma$ might be related). However, we will not investigate this here.\\

The second way to obtain algebraic relations involving the forms is to use the algebraic Killing spinor equation. As we will see, this relates combinations of forms and field strengths. The algebraic Killing spinor equation is given by $\delta \lambda = {\mathcal P} \eps = 0\ $ where \cite{Bergshoeff:1999bx}
\be
{\mathcal P} = \Gamma^A \partial_A \phi + \frac{1}{12} H_{A_1 A_2 A_3}\Gamma^{A_1 A_2 A_3}\otimes \sigma^3  + \frac{e^{\phi}}{4}\sum_{n=1}^5 \frac{(-1)^{n-1} (n-3)}{(2n-1)!} G_{A_1 \ldots A_{2n-1}}\Gamma^{A_1 \ldots A_{2n-1}}\otimes \lambda_n 
\ee
where the $2\times 2$ matrices $\lambda_n$ are given in Eq.~(\ref{lambdan}). Algebraic identities can be obtained from this equation by constructing $\bar\eps^i \Gamma_{M_1 \ldots M_p} ({\mathcal P}\eps)^j = 0$, for $p=0,1,\ldots, 10$. For $p=0\ $ we obtain the following set of identities,
\begin{eqnarray}
\bar\eps^1 \left({\mathcal P} \eps \right)^1 &=& K^{11}\cdot d \phi - e^{\phi} K^{12}\cdot G^{(1)} + \frac{e^{\phi}}{2} G^{(3)}\cdot \Phi^{12} = 0 \label{firstalg}\\
\bar\eps^2 \left({\mathcal P} \eps \right)^1 &=& K^{21} \cdot d\phi - e^{\phi} K^{22} \cdot G^{(1)} + \frac{1}{2} H\cdot \Phi^{21} = 0\\
\bar\eps^1 \left({\mathcal P} \eps \right)^2 &=& K^{12} \cdot d\phi + e^{\phi} K^{11}\cdot G^{(1)} - \frac{1}{2} H\cdot \Phi^{12} = 0\\
\bar\eps^2 \left({\mathcal P} \eps \right)^2 &=& K^{22}\cdot d\phi + e^{\phi} K^{21}\cdot G^{(1)} + \frac{e^{\phi}}{2} G^{(3)}\cdot \Phi^{21} = 0
\end{eqnarray}
The case $p=1$ gives no identities, but for $p=2$ we obtain another set of four identities given as follows,
\begin{eqnarray}
0 = \bar\eps^1 \Gamma_{NP} ({\mathcal P}\eps)^1 &=& \left( K^{11}\wedge d\phi - e^{\phi} K^{12}\wedge G^{(1)}  - e^{\phi} \iota_{G^{(1)}} \Phi^{12}\right)_{NP} + \frac{e^{\phi}}{2} \Phi^{12}_{A_1 A_2 [N}{G^{(3)}_{P]}}^{A_1 A_2}\nonumber\\
&& + \frac{1}{2}\left( \iota_H \Sigma^{11} - \iota_{K^{11}} H + e^{\phi}\iota_{G^{(3)}}\Sigma^{12} - e^{\phi} \iota_{K^{12}} G^{(3)}\right)_{NP}\\
&&\nonumber\\
0 = \bar\eps^2 \Gamma_{NP} ({\mathcal P}\eps)^1 &=& \left( K^{21}\wedge d\phi - e^{\phi}K^{22}\wedge G^{(1)} +\iota_{d\phi} \Phi^{21}\right)_{NP}+ \frac{1}{2}\Phi^{21}_{A_1 A_2 [N}{H^{}_{P]}}^{A_1 A_2}\nonumber\\
&& + \frac{1}{2}\left( \iota_H \Sigma^{21} - \iota_{K^{21}} H + e^{\phi}\iota_{G^{(3)}}\Sigma^{22} - e^{\phi}\iota_{K^{22}}G^{(3)}\right)_{NP}\\
&&\nonumber\\
0 = \bar\eps^1 \Gamma_{NP} ({\mathcal P}\eps)^2 &=& \left( K^{12}\wedge d\phi + e^{\phi}K^{11}\wedge G^{(1)}+\iota_{d\phi} \Phi^{12}\right)_{NP} - \frac{1}{2}\Phi^{12}_{A_1 A_2 [N}{H^{}_{P]}}^{A_1 A_2}\nonumber\\
&& + \frac{1}{2}\left( -\iota_H \Sigma^{12} + \iota_{K^{12}} H + e^{\phi}\iota_{G^{(3)}}\Sigma^{11} - e^{\phi}\iota_{K^{11}}G^{(3)}\right)_{NP}\\
&&\nonumber\\
0= \bar\eps^2 \Gamma_{NP} ({\mathcal P}\eps)^2 &=& \left( K^{22}\wedge d\phi + e^{\phi} K^{21}\wedge G^{(1)}  + e^{\phi} \iota_{G^{(1)}} \Phi^{21}\right)_{NP} + \frac{e^{\phi}}{2} \Phi^{21}_{A_1 A_2 [N}{G^{(3)}_{P]}}^{A_1 A_2}\nonumber\\
&& + \frac{1}{2}\left( -\iota_H \Sigma^{22} + \iota_{K^{22}} H + e^{\phi}\iota_{G^{(3)}}\Sigma^{21} - e^{\phi} \iota_{K^{21}} G^{(3)}\right)_{NP}
\end{eqnarray}
The final set of four identities comes from $p=4$. For example,
\begin{eqnarray}
0 = \bar\eps^1 \Gamma_{NPQR} ({\mathcal P}\eps)^1 &=& \left(\iota_{d\phi}\Sigma^{11} - e^{\phi} \iota_{G^{(1)}} \Sigma^{12}+ e^{\phi} G^{(1)}\wedge \Phi^{12}\right)_{NPQR}\nonumber\\
&& -\frac{1}{2} \left( K^{11} \wedge H + e^{\phi} K^{12} \wedge G^{(3)} - e^{\phi} \iota_{G^{(3)}}\Pi^{12}\right)_{NPQR}\nonumber\\ 
&& - {H^{A_1 A_2}}^{}_{[N}\Sigma^{11}_{PQR]A_1 A_2} - e^{\phi} G^{(3)}_{A_1 A_2 [N}{\Sigma^{12}_{PQR]}}^{A_1 A_2}   
\end{eqnarray}
Again, there are three other similar identities for $p=4$, given by
\begin{eqnarray}
0 = \bar\eps^2 \Gamma_{NPQR} ({\mathcal P}\eps)^1 &=& \left( \iota_{d\phi} \Sigma^{21} - e^{\phi}\iota_{G^{(1)}}\Sigma^{22}- d\phi\wedge \Phi^{21}\right)_{NPQR}\nonumber\\
&& -\frac{1}{2}\left( K^{21}\wedge H + e^{\phi} K^{22}\wedge G^{(3)} - \iota_{H}\Pi^{21}\right)_{NPQR}\nonumber\\
&& - {H^{A_1 A_2}}^{}_{[N}\Sigma^{21}_{PQR]A_1 A_2} - e^{\phi}G^{(3)}_{A_1 A_2 [N}{\Sigma^{22}_{PQR]}}^{A_1 A_2}\\
&&\nonumber\\
0 = \bar\eps^1 \Gamma_{NPQR} ({\mathcal P}\eps)^2 &=& \left( \iota_{d\phi} \Sigma^{12} + e^{\phi}\iota_{G^{(1)}}\Sigma^{11}- d\phi\wedge \Phi^{12}\right)_{NPQR}\nonumber\\
&& +\frac{1}{2}\left(  K^{12}\wedge H - e^{\phi} K^{11}\wedge G^{(3)} - \iota_{H}\Pi^{12}\right)_{NPQR}\nonumber\\
&& + {H^{A_1 A_2}}^{}_{[N}\Sigma^{12}_{PQR]A_1 A_2} - e^{\phi}G^{(3)}_{A_1 A_2 [N}{\Sigma^{11}_{PQR]}}^{A_1 A_2}\\
&&\nonumber \\
0 = \bar\eps^2 \Gamma_{NPQR} ({\mathcal P}\eps)^2 &=& \left(\iota_{d\phi}\Sigma^{22} + e^{\phi} \iota_{G^{(1)}} \Sigma^{21}- e^{\phi} G^{(1)}\wedge \Phi^{21}\right)_{NPQR}\nonumber\\
&&  +\frac{1}{2} \left(  K^{22} \wedge H - e^{\phi} K^{21} \wedge G^{(3)} + e^{\phi} \iota_{G^{(3)}}\Pi^{21}\right)_{NPQR}\nonumber\\
&& + {H^{A_1 A_2}}^{}_{[N}\Sigma^{22}_{PQR]A_1 A_2} - e^{\phi} G^{(3)}_{A_1 A_2[N}{\Sigma^{21}_{PQR]}}^{A_1 A_2}  \label{lastalg}
\end{eqnarray}
If we take $p>4$ in $\bar\eps^i \Gamma_{M_1 \ldots M_p} ({\mathcal P}\eps)^j = 0$, we obtain identities which are simply the duals of those obtained for $p<4$. Therefore, Eqs.~(\ref{firstalg})-(\ref{lastalg}) give the full set of independent identities that can be derived from the algebraic Killing spinor equation.


\section{Giant gravitons in $AdS_5\times S^5$ from holomorphic surfaces}
In this section we review the Mikhailov construction of giant gravitons in $AdS_5\times S^5$ via holomorphic surfaces \cite{Mikhailov:2000ya}. This construction will give rise to a simple set of supersymmetry projection conditions for a giant graviton probe. These projection conditions will allow the forms $K$, $\Phi$ and $\Sigma$ to be found and we will see in \S~5 that the differential equations satisfied by the forms correspond to calibration conditions for these branes.

\subsection{The complex structure of $AdS_5\times S^5$}\label{complex}

We begin by embedding the $S^5$ part of the geometry in flat $\mathbb{C}^3$, which has complex coordinates $Z_i$ ($i=1,2,3$), which can be written in terms of 6 real polar coordinates $\{ \mu_i, \phi_i\}$, $0\leq \phi_i \leq 2\pi$, as $Z_i = \mu_i e^{i\phi_i}$. The metric on $\mathbb{C}^3$ is given by
\be
ds^2 = |dZ_1|^2  + |dZ_2|^2 +|dZ_3|^2 = \sum_{i=1}^3 \left(d\mu_i^2 + \mu_i^2 d\phi_i^2\right) \label{C3}
\ee
$\mathbb{C}^3$ has a complex structure, $I$, which acts on the basis 1-forms as follows,
\[
I: dZ_i \longrightarrow - i dZ_i
\]
This is equivalent to the following transformations of the real 1-forms: $d\mu_i \longrightarrow \mu_i d\phi_i$ and $\mu_i d\phi_i\longrightarrow - d\mu_i$.
The sphere is defined in $\mathbb{C}^3$ by
\be
S^5 : |Z_1|^2 + |Z_2|^2 + |Z_3|^2 = \mu_1^2 + \mu_2^2 + \mu_3^2 = 1 \label{sphere}
\ee
where we have set the radius to 1 for convenience. Note that this means that the radius of curvature of $AdS_5$ is also 1.
The metric on $S^5$ is given by the metric on $\mathbb{C}^3$, Eq.~(\ref{C3}), restricted to the sphere.
The embedding of $S^5$ in $\mathbb{C}^3$ allows us to define a radial 1-form, $e^{r}\in T^* \mathbb{C}^3$, which is orthogonal to the sphere at every point. Explicitly, $e^{r}$ is given by
\[
e^r = \mu_1 d\mu_1 + \mu_2 d\mu_2 + \mu_3 d\mu_3
\]
We can act with the complex structure on $e^{r}$ to produce a new 1-form $e^{||} = I\cdot e^{r}$, which is given explicitly by
\be
e^{||} = \mu_1^2 d\phi_1 + \mu_2^2 d\phi_2 + \mu_3^2 d\phi_3 \label{e||}
\ee
On the sphere $e^{||}$ has unit length and it belongs to $T^* S^5$. Therefore, $e^{||}$ gives a preferred direction on the sphere. We will see later that this is the direction of motion for giant gravitons in this construction. Note that
\be
de^{||} = 2 (\mu_1 d\mu_1\wedge d\phi_1 + \mu_2 d\mu_2\wedge d\phi_2 + \mu_3 d\mu_3\wedge d\phi_3 ) \equiv 2 \w \label{w1}
\ee
where $\omega$ is the K\"{a}hler 2-form on $\mathbb{C}^3$. We can also write $\w$ in another basis as follows,
\be
\omega = {\mathcal N}e^{r}\wedge e^{||} + e^{I^1} \wedge e^{J^1} +  e^{I^2} \wedge e^{J^2} \label{w2}
\ee
Here $\{ e^{I^1},e^{J^1}, e^{I^2}, e^{J^2} \}$ are unit 1-forms on $\mathbb{C}^3$, where
\[ e^{J^k} = I \cdot e^{I^k}, \qquad k=1,2\]
These 1-forms are orthogonal to each other and to $\{e^{||}, e^{r}\}$. The factor of ${\mathcal N}$ in Eq.~(\ref{w2}) ensures that $e^r$ and $e^{||}$ are normalised everywhere on $\mathbb{C}^3$. Explicitly, ${\mathcal N} = (\mu_1^2 + \mu_2^2 + \mu_3^2)^{-1}$.
Since $e^{I}$ and $e^{J}$ are non-zero 1-forms on $S^5$, the restriction of $\w$ to the sphere is simply
\be
\omega{\Big{|}}_{S^5} = \left(e^{I^1} \wedge e^{J^1} +  e^{I^2} \wedge e^{J^2}\right){\Big{|}}_{S^5} \label{ws}
\ee
This restricted K\"{a}hler 2-form will be important later when we construct $p$-forms relevant to supersymmetric giant gravitons.\\

It is also possible to define a complex structure for $AdS_5$. In particular, we embed $AdS_5$ in flat $\mathbb{C}^{1,2}$, which has complex coordinates $W_a=u_a+ i v_a$ ($a=0,1,2$). The flat metric on $\mathbb{C}^{1,2}$ is given by
\begin{eqnarray*}
ds^2 &=& - |dW_0|^2  + |dW_1|^2 +|dW_2|^2
\end{eqnarray*}
$\mathbb{C}^{1,2}$ also has a complex structure, $\tilde{I}$, which acts on the basis 1-forms as 
\[
\tilde{I}: dW_a \longrightarrow - i dW_a
\]
i.e. $du_a \longrightarrow dv_a$ and $dv_a \longrightarrow - du_a$.
The embedding of $AdS_5$ in $\mathbb{C}^{1,2}$ is given by
\be
|W_0|^2 - |W_1|^2 - |W_2|^2 = 1 \label{AdS5}
\ee
The metric on $AdS_5$ is given by the metric on $\mathbb{C}^{1,2}$ restricted to this surface. 
In a similar way to the $S^5$, we can define a radial 1-form, $e^{\perp}$, which is orthogonal to $AdS_5$ at every point. Explicitly, $e^{\perp}$ is given by
\[
e^{\perp} = u_0 du_0 + v_0 dv_0 - u_1 du_1 - v_1 dv_1 - u_2 du_2 - v_2 dv_2
\]
We can act with the complex structure on $e^{\perp}$ to obtain a time-like direction, $e^0 = \tilde{I}\cdot e^{\perp}$, which belongs to the cotangent space of $AdS_5$:
\be
e^0 =  u_0 dv_0 - v_0 du_0 - u_1 dv_1 + v_1 du_1 - u_2 dv_2 + v_2 du_2
\ee
This is a preferred timelike direction on $AdS_5$ which will be used later in the supersymmetry projection conditions for giant gravitons. The derivative of $e^0$ is related to the K\"{a}hler form on $\mathbb{C}^{1,2}$, denoted $\tilde{\w}$, by
\be
d e^0 = 2(  du_0\wedge dv_0 - du_1\wedge dv_1 -  du_2\wedge dv_2 ) \equiv - 2 \tilde{\w} \label{tildew11}
\ee
In a local region close to the sphere (so that $e^{\perp}$ and $e^0$ remain time-like), $\tilde{\w}$ can also be written in a different basis as
\be
\tilde{\w} = - \tilde{\mathcal N} e^{\perp}\wedge e^0 + e^{a_1} \wedge e^{b_1} + e^{a_2} \wedge e^{b_2} \label{tildew1}
\ee
where $e^{b_k} = \tilde{I}\cdot e^{a_k}$, $k=1,2$, are unit spacelike 1-forms on $\mathbb{C}^{1,2}$ and $\tilde{\mathcal N}$ normalises $e^{\perp}$ and $e^0$ in this region.
The above form for $\tilde{\w}$ restricts conveniently to $AdS_5$ as
\be
\tilde{\w}{\Big{|}}_{AdS} = (e^{a_1} \wedge e^{b_1} + e^{a_2} \wedge e^{b_2}){\Big{|}}_{AdS} \label{wads}
\ee
This 2-form will appear later in the calibrating form for giant gravitons.\\

Later it will be useful to parameterise $AdS_5$ with ``polar'' coordinates. In particular, we can take
\[ W_0 = \cosh\rho\ e^{it},\qquad W_1 = \sinh\rho\ (\Omega_1 + i \Omega_2), \qquad W_2 = \sinh\rho\ (\Omega_3+ i \Omega_4)
\]
where $\sum_{i=1}^4 \Omega_i^2 = 1$. With these definitions the embedding condition for $AdS_5$, given in Eq.~(\ref{AdS5}), is automatically satisfied. In these coordinates, the metric on $AdS_5$ is given by
\be
ds_{AdS}^2 = -\cosh^2\rho\ dt^2 + d\rho^2 + \sinh^2\rho\ \sum_{i=1}^4 d\Omega_i^2 \label{AdSmetric}
\ee
supplemented with the condition that $\sum_{i=1}^4 \Omega_i^2 = 1$. The timelike 1-form, $e^0$, is
\be
e^0 =  \cosh^2\rho\ dt - \sinh^2\rho\ (\Omega_1 d\Omega_2 - \Omega_2 d\Omega_1 + \Omega_3 d\Omega_4 - \Omega_4 d\Omega_3) \label{e0exp}
\ee

\subsection{Giant graviton construction}\label{construction}
Giant gravitons in $AdS_5\times S^5$ are D3-branes which have their spatial world-volume entirely contained in the $S^5$ part of the geometry. In this construction, the spatial world-volume of the brane is defined by the intersection of a holomorphic surface in $\mathbb{C}^3$ with the $S^5$. In particular, we consider the class of holomorphic surfaces, $C\subset\mathbb{C}^3$, which have complex dimension $2$ (4 real dimensions). These surfaces are specified by a single equation,
\[F(Z_1, Z_2, Z_3)=0\] 
Here $F$ depends only on the holomorphic coordinates $Z_i$ (but not the $\bar{Z}_i s$). The intersection of $C$ with $S^5$ is a 3-dimensional surface, $\Sigma$, which we take to be the spatial world-volume of the giant graviton at time $t=0$.\\

Giant gravitons have a non-trivial motion on the $S^5$. In this construction they are defined to move with the speed of light (c=1 in our units) in the direction $e^{||}$. Typically the surface of the giant graviton $\Sigma$ will not be orthogonal to $e^{||}$ (in fact the construction would break down if the brane was completely orthogonal to $e^{||}$ at any point). Therefore, at each point on the brane $e^{||}$ can be decomposed into a component normal to the brane, denoted $e^{\phi}$, and a component parallel to the brane, denoted $e^{\psi}$, i.e.
\be
e^{||} = - v e^{\phi} - \sqrt{1- v^2}e^{\psi} \label{decomp}
\ee
where $0<v<1$. In fact, $v$ turns out to be the speed of the giant graviton in the direction $e^{\phi}$. This association arises from requiring the brane to be supersymmetric \cite{Mikhailov:2000ya}, and we will see in \S~\ref{calgiants} that this condition is also encoded in the calibration bound for giant gravitons. Since $v<1$, it means that the surface elements of the brane move at less than the speed of light, even though the centre of mass of the brane (which does not lie on the brane) moves with the speed of light.\\

We can actually define the full world-volume of the giant graviton using the holomorphic function $F$. Due to the form of $e^{||}$, given in Eq.~(\ref{e||}), the full world-volume of the giant graviton is given by the intersection of $S^5$ with the following surface \cite{Mikhailov:2000ya}
\[
F(e^{it} Z_1, e^{it} Z_2,e^{it} Z_3) = 0 
\]
The above equation describes the original holomorphic surface translated in the direction $e^{||}$ at the speed of light. The usual giant graviton, which was first discussed in Ref.~\cite{McGreevy:2000cw}, is a simple case of this construction; one takes the holomorphic surface to be simply $F= Z_1 - c$, where $c$ is a constant. 
However, more complicated giant gravitons are also included in this description, since any holomorphic surface can be used. 
Mikhailov proves that all giant gravitons in this construction preserve at least $\frac{1}{8}$ supersymmetry.
We now discuss the supersymmetry projection conditions for these objects. 

\subsection{Giant gravitons and supersymmetry}\label{susy}

This construction of giant gravitons via holomorphic surfaces in the 12-dimensional complex space $\mathbb{C}^{1,2}\times \mathbb{C}^3$ means that they preserve supersymmetry (provided $v$ is associated with the physical speed of the giant graviton, see Ref.~\cite{Mikhailov:2000ya} for details). Moreover, the supersymmetry projection conditions can be written down in a very simple way. This is essentially because Killing spinors in $AdS_5\times S^5$ become covariantly constant spinors in the 12-dimensional space, so everything simplifies in the higher-dimensional setting. The amount of supersymmetry preserved by a particular giant graviton depends on the function $F$ which defines the holomorphic surface $C$ (and hence the brane surface $\Sigma$). If $F$ depends on (1, 2, 3) of the complex coordinates\footnote{Up to linear holomorphic redefinitions of $Z_i$ which do not alter the amount of supersymmetry preserved.}
then the resulting giant graviton configuration will preserve ($\frac{1}{2}$, $\frac{1}{4}$, $\frac{1}{8}$) of the supersymmetry respectively. 
The projection conditions satisfied by the most general configurations, which preserve $\frac{1}{8}$ supersymmetry, are given by \cite{Mikhailov:2000ya}
\begin{eqnarray}
\Gamma^0 \Gamma^{||} \eps^1 &=& \eps^1 \nonumber\\
\Gamma^{I^k}\Gamma^{J^k}\eps^1 &=& -\eps^2, \qquad k=1,2 \label{proj2}
\end{eqnarray}
Here $\eps^1, \eps^2$ are the 16-component spinors associated to a 32-dimensional Killing spinor, $\eps$, of $AdS_5\times S^5$. Moreover,
\[\Gamma^{||} = \Gamma(e^{||}), \qquad \Gamma^0 = \Gamma(e^0), \qquad \Gamma^{I^k} = \Gamma(e^{I^k}), \qquad \Gamma^{J^k} = \Gamma(e^{J^k}) \]
where the 1-forms $e^{||}, e^0, e^{I^k}, e^{J^k}$ are defined in \S~\ref{complex} and here they are all evaluated on $AdS_5\times S^5$ (N.B. while these projections are made with reference to the complex structure of $\mathbb{C}^{1,2}\times \mathbb{C}^3$, everything now is in 10 dimensions, so the forms must be evaluated on the lower-dimensional space). The projection conditions in Eq.~(\ref{proj2}) are manifestly brane independent. We will use these projection conditions to explicitly construct the differential forms, $K^{ij}$, $\Phi^{ij}$ and $\Sigma^{ij}$, which were defined in \S~\ref{pforms}. We will see in \S~5 that the 3-form, $\Phi^{ij}$, can be interpreted as a calibrating form for giant gravitons.

\subsection{The differential forms for giant gravitons}\label{diffgiants}
To explicitly work out the forms, we need to make some additional projections which are compatible with Eq.~(\ref{proj2}), to ensure that there is only one independent Killing spinor, $\eps$, which satisfies the conditions. The projection conditions in Eq.~(\ref{proj2}) admit 4 independent Killing spinors, so we need to make another two projections to reduce this number to 1  (because each projection reduces the number of allowed spinors by $\frac{1}{2}$).
The obvious way to make compatible projections is to treat the complex structure of $AdS_5$ in a similar way to the complex structure of $S^5$. 
Therefore, one set of possible projections is
\be
\Gamma^{a_k}\Gamma^{b_k} \eps^1 = -\eps^2 \qquad k=1,2 \label{proj3}
\ee
where 
\[\Gamma^{a_k} = \Gamma(e^{a_k}), \qquad \Gamma^{b_k} = \Gamma(e^{b_k})\]
and $e^{a_k}, e^{b_k}$, defined in \S~\ref{complex}, are non-zero 1-forms which we evaluate on $AdS_5$. These 1-forms are orthogonal to $\{ e^{0}, e^{||}, e^{I^k}, e^{J^k}\}$, so the above projections commute with the existing projections in Eq.~(\ref{proj2}). Therefore, the full set of projection conditions consists of Eqs.~(\ref{proj2}) and (\ref{proj3}). Note that in this basis the chirality condition, $\Ga^{0\ldots 9} \eps^i = \eps^i$, becomes
\[
\Ga^{ 0 a_1 b_1 a_2 b_2 I^1 J^1 I^2 J^2 ||}\eps^i = \eps^i
\]\\

Using the projection conditions given in Eqs.~(\ref{proj2})-(\ref{proj3}) we can now compute all the differential forms which were defined in \S~\ref{pforms}. This will give us the set of $p$-forms relevant to giant gravitons. Firstly, the 1-forms, $K^{ij}$, are given by
\begin{eqnarray}
K^{11} &=& K^{22} = \Delta(e^0 + e^{||}) \label{K11gg}\\
K^{12} &=& K^{21} = 0\nonumber
\end{eqnarray}
where $\Delta$ is the normalisation of the spinors $(\eps^1)^t \eps^1 = (\eps^2)^t \eps^2 = \Delta$. The 3-forms are given by
\be
\Phi^{12} = - \Phi^{21} = \Delta (e^0 + e^{||})\wedge (\w_S + \tilde{\w}_{AdS}) \label{phi12gg}
\ee
where $\w_S$ and $\tilde\w_{AdS}$ are defined in Eqs.~(\ref{ws}) and (\ref{wads}) respectively. The 5-forms are given by
\begin{eqnarray}
\Sigma^{11} &=& \Sigma^{22} = \Delta (e^0 + e^{||})\wedge \left( - \frac{1}{2}\w_S\wedge \w_S  - \frac{1}{2}\tilde\w_{AdS}\wedge\tilde\w_{AdS} - \w_S \wedge \tilde\w_{AdS}\right)\label{Sigma}\\
\Sigma^{12} &=& \Sigma^{21} = 0 . \nonumber
\end{eqnarray}
The expressions for the 7-forms $\Pi^{ij}$ and the 9-forms $\Omega^{ij}$ can be found by dualizing the relevant lower-degree forms.\\ 

We now calculate the derivatives of all these forms and show that they obey the differential equations derived in \S~\ref{pforms}. The derivatives of the forms will be related to $G^{(5)}$, which is the only non-zero field strength in $AdS_5\times S^5$. Now $G^{(5)} = -4 \{vol(AdS_5) + vol(S^5)\}$. In our basis this is\be
G^{(5)} = - 2 (e^0 \wedge \tilde\w_{AdS} \wedge \tilde\w_{AdS} + e^{||}\wedge \w_S \wedge \w_S)
\ee
To calculate the derivatives of the forms we will need the following results,
\begin{eqnarray}
de^0 &=& - 2 \tilde\w_{AdS}, \nonumber\\
de^{||} &=& 2\w_S, \nonumber\\
d\Delta &=&0 \label{3eqs}
\end{eqnarray}
The first two equations follow from Eqs.~(\ref{w1}) and (\ref{tildew11}), together with the fact that the 1-forms $e^0$ and $e^{||}$ are evaluated on $AdS_5\times S^5$. The third equation can be derived by writing $\Delta = \bar\eps^1 \Gamma_0 \eps^1$ and calculating $d\Delta$ using the Killing spinor equation. We will not go into the details of this calculation here, but it is straight-forward. Note that the third equation allows us to set $\Delta = 1$, which we do in the following.\\


We first consider the differential equation for $K^{11}$. From Eqs.~(\ref{K11gg}) and (\ref{3eqs}) the derivative is given by
\be
d K^{11} = d e^0 + d e^{||} = 2 (- \tilde\w_{AdS} + \w_S)
\ee
From Eq.~(\ref{K11}), this should be related to $\iota_{\Phi^{12}} G^{(5)}$, which we now compute:
\[
\iota_{\Phi^{12}} G^{(5)}  = -  4 (- \tilde\w_{AdS} + \w_S)
\]
Therefore, 
\[
d K^{11} =  - \frac{1}{2} \iota_{\Phi^{12}} G^{(5)}
\]
This is precisely what we expect from Eq.~(\ref{K11}) because the dilaton is constant for $AdS_5 \times S^5$. The equation for $K^{22}$ works in exactly the same way, and the equations for $K^{12}$ and $K^{21}$ are trivially satisfied as both left and right hand sides of the equations are identically zero.\\

We now consider the differential equation for $\Phi^{12}$. Firstly, from Eqs.~(\ref{phi12gg}) and (\ref{3eqs}) we have
\begin{eqnarray}
d\Phi^{12} &=& d \left[ (e^0 + e^{||}) \wedge (\w_S + \tilde\w_{AdS})\right] \nonumber\\
&=& 2 (\w_S\wedge \w_S - \tilde\w_{AdS}\wedge \tilde\w_{AdS}) \label{dphi}
\end{eqnarray}
From Eq.~(\ref{phi12}) this should be related to $\iota_{(K^{11} + K^{22})}G^{(5)}$ which we can compute:
\be
\iota_{(K^{11} + K^{22})}G^{(5)} = 4 (\tilde\w_{AdS}\wedge\tilde\w_{AdS} - \w_S\wedge \w_S) \label{iKG}
\ee
Therefore, from Eqs.~(\ref{dphi}) and (\ref{iKG}), we have
\be
d\Phi^{12} = - \frac{1}{2} \iota_{(K^{11} + K^{22})}G^{(5)} \label{phieq}
\ee
as required. The equation for $\Phi^{21}$ works in the same way since $\Phi^{12} = - \Phi^{21}$. Note that Eq.~(\ref{phieq}) is very similar to the condition for a generalised calibration \cite{Gutowski}. In the next section we will see precisely how $\Phi$ is related to a generalised calibration for giant gravitons.\\

From Eqs.~(\ref{Sigma}) and (\ref{3eqs}), the derivative of the 5-form is
\be
d \Sigma^{11} = d\Sigma^{22} = - \tilde\w_{AdS} \wedge \w_S\wedge \w_S + \w_S \wedge \tilde\w_{AdS}\wedge \tilde\w_{AdS}
\ee
From Eq.~(\ref{Sigma11}) we have that the components $(d\Sigma^{11})_{MNPQRS}\ $ should be equal to 
\[
\frac{15}{2} { \Phi^{12}}_{A[MN}{G^{(5)}_{PQRS]}}^A
\]
since $K^{12}=K^{21}=0\ $. By considering different combinations of the indices, one finds that 
\[ \frac{15}{2} { \Phi^{12}}_{A[MN}{G^{(5)}_{PQRS]}}^A = {\left(- \w_S\wedge \w_S\wedge \tilde\w_{AdS} + \w_S \wedge \tilde\w_{AdS} \wedge \tilde\w_{AdS}\right)}_{MNPQRS} 
\]
and hence
\be
d\Sigma^{11} = \frac{15}{2} { \Phi^{12}}_{A[MN}{G^{(5)}_{PQRS]}}^A
\ee 
as required. The equation for $\Sigma^{22}$ works in the same way, and the equations for $\Sigma^{12}$ and $\Sigma^{21}$ are trivially satisfied.

\section{Calibrations for giant gravitons}\label{calgiants}

In this section we will show that the differential forms constructed in the previous section can be used as calibrating forms for giant gravitons. We will see that all giant gravitons constructed from holomorphic surfaces are calibrated. To begin we consider the super-translation algebra for D3-branes in type IIB supersymmetric backgrounds. This algebra will allow us to find a calibration bound for giant gravitons, and we will see that the bound involves the 3-forms $\Phi^{ij}$.

\subsection{The super-translation algebra}

The super-translation algebra for D3-branes in flat space is given by \cite{KH}
\be
\{Q_{i\al},Q_{j\beta }\} = \delta_{ij} (C \Gamma_M)_{\al\beta} {P}^M  + (i\sigma_2)_{ij} (C \Gamma_{MNP})_{\al\beta} Z^{MNP} \label{supertrans}
\ee
where
\be
Z^{MNP} = \frac{1}{3!}\int dX^M\wedge dX^N\wedge dX^P \label{Z}
\ee
and the integral is taken over the spatial world-volume of the brane. 
The indices $i, j \in \{1,2\}$ label the 16-dimensional spinors and $\al,\beta$ are spinor indices. The matrix $C$ is the charge conjugation matrix, which we will take to be $\Gamma^0$ from now on. The quantity $P^M$ is the total 10-momentum of the brane. The term involving $Z$ is a topological charge for the D3-brane. The fact that it is topological is clear from Eq.~(\ref{Z}) since $Z$ is defined as the integral of a {\it closed form} over the spatial world-volume of the brane. We now introduce a constant 32-dimensional spinor, $\eps = \eps^{i\al}$, and contract all indices in the super-translation algebra, Eq.~(\ref{supertrans}), with the indices of $\eps$ to obtain
\be
2 (Q\eps)^2 = (K^{11} + K^{22})\cdot P + \int (\Phi^{12} - \Phi^{21}) \label{intermediate}
\ee
where 
\[
Q\eps = \sum_{i, \al} Q_{i\al} \eps^{i\al}
\] 
and $K^{11} = \bar\eps^1 \Gamma\eps^1$ etc. are the previously defined forms. We can also rewrite the first term in Eq.~(\ref{intermediate}) as an integral over the spatial world-volume of the brane as follows
\be
2 (Q\eps)^2 =  \int (K^{11} + K^{22})\cdot p + \int 2 \Phi^{12} \label{flatalg}
\ee
where $p_M$ is the momentum density on the brane world-volume. At the moment we are still considering flat space, so all background field strengths are zero, and hence from Eq.~(\ref{phi12}) the integrand $\Phi^{12}$ is closed. Therefore, the brane charge term is topological, as required.\\

We now want to consider the super-translation algebra for a curved background with non-zero $G^{(5)}$, but with all other field strengths zero. This will allow us to consider the case we are interested in, namely D3-brane giant gravitons in $AdS_5\times S^5$. Following Ref.~\cite{Hackett-Jones:2003vz} we can find the curved space super-translation algebra by modifying Eq.~(\ref{flatalg}) as follows. First we promote the constant spinor $\eps$ to a Killing spinor of the background. This means that the forms $K^{11}, K^{22}, \Phi^{12}$ are no longer constant, but become fields.
Secondly, we replace $\Phi^{12}$ by a closed 3-form, since for non-zero $G^{(5)}$,
\be
d \Phi^{12} = - \frac{1}{2} \iota_{K^{11} + K^{22}} G^{(5)} \label{phithing}
\ee
i.e. the integrand in Eq.~(\ref{flatalg}) is not closed. However, we can construct a closed 3-form from $\Phi^{12}$ by manipulating this equation. The starting point is to compute the Lie derivative of $G^{(5)}$ along the direction $K \equiv K^{11} + K^{22}$. In general, the Lie derivative of a $p$-form, $\w$, along a vector field, $X$, is
\be
\lal_X \w = d (\iota_X \w) + \iota_X d\w
\ee
Therefore,
\[
\lal_{K} G^{(5)} =d (\iota_{K} G^{(5)}) + \iota_{K} d G^{(5)} 
\]
Using Eq.~(\ref{phithing}) and the fact that $d G^{(5)}=0$, it is easy to see that the two terms here vanish independently and $\lal_{K} G^{(5)} = 0$. This means we can choose a gauge for the 4-form Ramond-Ramond potential $C^{(4)}$ (which is related to the 5-form field strength by $G^{(5)}= dC^{(4)}$) such that $\lal_{K} C^{(4)} = 0$ also. In that case
\[
d(2 \Phi^{12} - \iota_{K} C^{(4)} )= - \iota_{K} G^{(5)} - d \iota_{K} C^{(4)}= - \lal_K C^{(4)}  = 0
\]
Therefore, we propose that the 3-form $2 \Phi^{12}$ should be replaced by
\[
2 \Phi^{12}  - \iota_{K} C
\]
in the super-translation algebra for backgrounds with non-zero 5-form field strength, i.e. the algebra becomes 
\be
2 (Q\eps)^2 =  \int K \cdot p + \int \left(2 \Phi^{12}  - \iota_K C\right)
\ee
Clearly this reduces to the original flat space algebra if we set the 4-form potential to zero. We now use the fact that $(Q\eps)^2 \geq 0$ to obtain the following calibration bound:
\be
\int \left( K \cdot p - \iota_K C\right) \geq - \int 2 \Phi^{12} \label{bound1}
\ee
where the integral is over the spatial world-volume of the brane. This bound is valid for all D3-branes in supersymmetric backgrounds which have field strengths $G^{(1)}$, $G^{(3)}$ and $H$ identically zero. In particular, we will see in \S~\ref{holgiants} that holomorphic giant gravitons saturate this bound, i.e. they are calibrated. Moreover, in \S~\ref{dualgiants} we will see that dual giant gravitons are also calibrated.\\
 
First, however, we show that any brane which saturates the bound Eq.~(\ref{bound1}) (i.e. is calibrated) minimises the quantity $\int K\cdot p\ $ in its homology class. To prove this we consider two 3-dimensional manifolds $U$ and $V$ in the same homology class. Moreover, we assume that the manifold $U$ is calibrated, i.e.
\be
\int_U \left(K \cdot p - \iota_K C\right) = - \int_U 2 \Phi^{12}
\ee 
Now since $U$ and $V$ are in the same homology class, we can write $U = V + \partial \Xi$ where $\partial \Xi$ is the boundary of a 4-dimensional manifold $\Xi$. Therefore,
\be
\int_U \left( K \cdot p - \iota_K C\right) = - \int_{V + \partial \Xi} 2 \Phi^{12} \label{start}
\ee
Now using Stoke's theorem together with Eq.~(\ref{phithing}) we have
\[
- \int_{V + \partial \Xi} 2 \Phi^{12} = - \int_V 2  \Phi^{12} + \int_{\Xi} \iota_K G^{(5)}
\]
Since we have chosen a gauge where $\lal_K C =0$, it follows that $\iota_K G^{(5)}= - d\iota_K C$, and therefore,
\[
\int_{\Xi} \iota_K G^{(5)} = - \int_{\partial \Xi} \iota_K C = - \int_U \iota_K C + \int_V \iota_K C
\]
where we have used Stoke's law again, and rewritten $\partial \Xi = U-V$ in the last step.
Therefore, Eq.~(\ref{start}) becomes
\begin{eqnarray}
\int_U \left( K \cdot p - \iota_K C\right)  &=& - \int_V 2 \Phi^{12} - \int_U \iota_K C + \int_V \iota_K C\nonumber\\
&\leq& \int_V \left( K \cdot p - \iota_K C\right) - \int_U \iota_K C + \int_V \iota_K C\nonumber
\end{eqnarray}
where we have used the calibration bound Eq.~(\ref{bound1}) to replace the first term. Rearranging this is just,
\be
\int_U  K \cdot p \leq \int_V  K \cdot p
\ee
i.e. $U$ has minimal $\int  K \cdot p\ $ in its homology class.\\ 

To get some idea of what this means, we can consider the case where $K$ is simply the timelike vector $e^0$. In this case, $ K\cdot p = -p_0\ $ and $-p_0$ can be identified with the Hamiltonian for the D3-brane \cite{Gutowski}. Therefore, the quantity minimised by calibrated manifolds is the energy. For giant gravitons, however, $K$ is a null vector. In this case, the quantity minimised by calibrated surfaces is ``Energy minus momentum'', as we now see.


\subsection{Holomorphic giant gravitons}\label{holgiants}

We now specialise to the case of giant gravitons. First we consider the calibration bound Eq.~(\ref{bound1}) with $K$ and $\Phi$ relevant to holomorphic giant gravitons. Using Eqs.~(\ref{K11gg}) and (\ref{phi12gg}) we obtain,
\be
\int \left(- p_0 + p_{||} - \iota_0 C - \iota_{||} C\right) \geq \int - (e^0 + e^{||})\wedge (\w_S + \tilde\w_{AdS})  \label{AdSbound}
\ee
where the integrals are over the spatial world-volume of the brane.
Since the spatial world-volume of a giant graviton is entirely contained in the $S^5$ part of the geometry, the bound becomes
\be
\int \left({\mathcal H} + p_{||} - \iota_0 C - \iota_{||} C\right) \geq \int - e^{||}\wedge \w_S  \label{bound}
\ee
where we have identified $-p_0$ with the Hamiltonian density ${\mathcal H}$. From the previous section we know that calibrated branes minimise $\int  K\cdot p$, which in this case is 
\[
\int  K \cdot p\  = 2 \int \left(- p_0 + p_{||}\right)\ \propto\ \int \left({\mathcal H} + p_{||}\right) 
\]
Now recall that the physical direction of motion of the giant graviton surface is $e^{\phi}$, where $e^{||} = - v e^{\phi} - \sqrt{1-v^2} e^{\psi}$. There is no physical momentum in the direction $e^{\psi}$, so the quantity minimized by a calibrated giant graviton is
\[
 \int \left({\mathcal H} + p_{||}\right) = \int  \left({\mathcal H} -v p_{\phi}\right)
\]
i.e. calibrated giant gravitons minimise the total energy minus the total physical momentum, $J = \int v p_{\phi}$, which is a conserved charge. Note that this agrees with Ref.~\cite{Arapoglu:2003ti} where the generator of time translations for giant gravitons is $E-J$ (N.B. our definition of the direction $\phi$ is different to the definition in Ref.~\cite{Arapoglu:2003ti}). We will now see that giant gravitons constructed from holomorphic surfaces saturate the bound in Eq.~(\ref{bound}) and hence have minimal energy minus momentum. Moreover, we will see that a brane which wraps the same surface as a holomorphic giant graviton, but travels at the wrong speed, does not saturate the bound.\\

We begin by evaluating the quantities ${\mathcal H}$ and $p_{||}$ which appear on the left hand side of the bound Eq.~(\ref{bound}). To do this we must first calculate the giant graviton Lagrangian. Schematically, this is given by
\be
\lal = - \sqrt{-g} + {\mathcal P}(C^{(4)})
\ee
where $g$ is the determinant of the induced metric on the brane, and ${\mathcal P}(C^{(4)})$ is the pull-back of the 4-form potential to the giant graviton world-volume. For simplicity, we will assume that all giant gravitons we consider lie at $\rho=0$ and at fixed $\Omega_i$ in the $AdS$ space. The Mikhailov construction does not specify the trajectory of the giant graviton in the $AdS$ space. However, we know that giant gravitons are free massive particles in $AdS$, so they travel along time-like geodesics \cite{Page:2002xz}. The trajectory $\rho=0$ is one particular time-like geodesic in $AdS$, and it can be related to any other time-like geodesic in $AdS$ by an appropriate change of coordinates \cite{Caldarelli:2004yk}.
To calculate the induced metric, we rewrite the metric on $S^5$ in a basis which is related to the giant graviton world-volume:
\be
ds_{S^5}^2 = (e^{\phi})^2 + (e^n)^2 + d\Sigma^2  \label{sphere2}
\ee
Here $e^{\phi}$, defined in Eq.~(\ref{decomp}), is the physical direction of motion of the brane, and $e^n$ is a unit 1-form on $S^5$ which is orthogonal to $e^{\phi}$ and to the brane surface, $\Sigma$. The 3-dimensional metric $d\Sigma^2$ is the metric on the spatial world-volume of the giant graviton. 
This rewriting allows us to calculate the induced metric very easily. We obtain,\be
ds_{g.g.}^2 = (-1 + \dot\phi^2) dt^2 + d\Sigma^2 \label{induced}
\ee
where $\dot\phi = d e^{\phi} /dt$ and $t$ is the time coordinate on the brane. Note that the term $-dt^2$ comes from the pull-back of $AdS_5$ metric to the trajectory $\rho =0$. Since the giant graviton moves in the $e^{\phi}$ direction, the quantity ${\mathcal P}(C^{(4)})$ is simply given by
\[
{\mathcal P}(C^{(4)}) = C_{t \sigma^1 \sigma^2 \sigma^3} + \dot\phi C_{\phi\ \sigma^1 \sigma^2 \sigma^3}
\]
where $\sigma^i$ ($i=1,2,3$) are the coordinates on the world-space of the brane. Therefore, we obtain the following Lagrangian for the giant graviton,
\be
\lal = - \sqrt{(1-\dot\phi^2) \Sigma} + C_{t \sigma^1 \sigma^2 \sigma^3} + \dot\phi C_{\phi\ \sigma^1 \sigma^2 \sigma^3} 
\ee
where $\Sigma$ is the determinant of the metric $d\Sigma^2$. From this Lagrangian we can calculate the momentum conjugate to $\phi$. We obtain,
\be
p_{\phi} = \frac{\partial \lal}{\partial \dot\phi} = \frac{\dot\phi\ \sqrt{\Sigma}}{\sqrt{1-{\dot\phi}^2}} + C_{\phi\ \sigma^1 \sigma^2 \sigma^3}
\ee
Therefore, the Hamiltonian is
\be
{\mathcal H} = p_{\phi} \dot\phi - \lal = \frac{\sqrt{\Sigma}}{\sqrt{1-{\dot\phi}^2}} - C_{t \sigma^1 \sigma^2 \sigma^3}
\ee
Using the change of basis given in Eq.~(\ref{decomp}) the momentum in the direction $e^{||}$ is $p_{||}= - v p_{\phi}$, i.e.
\[
p_{||} = - \frac{v \dot\phi\ \sqrt{\Sigma}}{\sqrt{1-{\dot\phi}^2}} - v C_{\phi\ \sigma^1 \sigma^2 \sigma^3}
\]
Moreover, 
\[
\int \left(\iota_0 C + \iota_{||} C\right) =  \int \left(- C_{t\sigma^1 \sigma^2 \sigma^3} -  v\ C_{\phi\sigma^1 \sigma^2 \sigma^3}\right)\ d^3\sigma
\]
where we have used Eq.~(\ref{decomp}) together with the fact that $e^0 = dt$ on the giant graviton trajectory.
So the left hand side of the calibration bound Eq.~(\ref{bound}) becomes
\be
\int \left({\mathcal H} + p_{||} - \iota_0 C - \iota_{||} C\right)  =  \int \frac{\sqrt{\Sigma} ( 1- v\dot\phi)}{\sqrt{1-\dot\phi^2}}\ d^3\sigma \label{LHS}
\ee
Note that in the Mikhailov construction $\dot\phi = v$. However, one could also consider a brane which wraps the same surface $\Sigma$, but has a different speed in the direction $e^{\phi}$, $\dot\phi \neq v$. These branes are not calibrated. To see this we leave $v$ and $\dot\phi$ as distinct quantities for the moment.\\

We now evaluate the right hand side of Eq.~(\ref{bound}) for our general giant graviton. Using Eq.~(\ref{decomp}) we have
\[
e^{||}\wedge \w_S = - \sqrt{1-v^2}\ e^{\psi} \wedge \w_S
\]
We now use the fact, proved in Ref.~\cite{Mikhailov:2000ya}, that the surface of the brane, $\Sigma$, wraps a 3-cycle consisting of the direct product of a 1-cycle, $e^{\psi}$, with a complex 2-cycle. Therefore, the pull-back of $e^{\psi} \wedge \w_S$ to the brane is simply the spatial world-volume of the brane, i.e.
\[
\int e^{\psi} \wedge \w_S  = \int \sqrt{\Sigma}\ d^3 \sigma 
\]
Therefore, the right hand side of Eq.~(\ref{bound}) is 
\be
\int - e^{||}\wedge \w_S= \int \sqrt{(1-v^2) \Sigma}\ d^3\sigma \label{RHS}
\ee
Clearly, the left and right hand sides of the calibration bound, Eqs.~(\ref{LHS}) and (\ref{RHS}), are equal when $\dot\phi =v$, which is the case for the giant graviton. This means that holomorphic giant gravitons are calibrated. For a giant graviton moving at the ``wrong speed'', i.e. $\dot\phi \neq v$, then
\[
\frac{\sqrt{\Sigma} ( 1- v\dot\phi)}{\sqrt{1-\dot\phi^2}} > \sqrt{(1-v^2) \Sigma}
\]
i.e. the brane is not calibrated, but it does satisfy the calibration bound in Eq.~(\ref{bound}).

\subsection{Dual giant gravitons}\label{dualgiants}

So far we have shown that giant gravitons constructed from holomorphic surfaces are calibrated. An interesting extension is to look at dual giant gravitons. Dual giants are D3-branes which wrap a 3-sphere in $AdS_5$. Like giant gravitons they have a non-trivial motion on the $S^5$ part of the geometry, but dual giants do not wrap any of the $S^5$ directions. 
In this section we show that the dual giant graviton introduced in Ref.~\cite{Grisaru:2000zn} saturates the bound Eq.~(\ref{bound1}). That is, we will show that
\[
\int \left(K\cdot p - \iota_K C\right)  = - \int 2 \Phi^{12}
\] 
for this configuration. 
Now it is known that dual giants preserve the same basic supersymmetries as giant gravitons \cite{Grisaru:2000zn}. Therefore, the forms $K$, $\Phi$ and $\Sigma$ will be exactly the same for the dual giants as for the ordinary giant gravitons. These forms are given explicitly in Eqs.~(\ref{K11gg})--(\ref{Sigma}). Therefore, the calibration bound reduces to
\be
\int \left({\mathcal H} + p_{||} - \iota_0 C- \iota_{||} C\right) \geq \int - (e^0 + e^{||}) \wedge (\w_S + \tilde\w_{AdS}) 
\ee
exactly as for giant gravitons. However, because dual giant gravitons wrap three $AdS$ directions, the only term on the right hand side that contributes is $- \int e^0 \wedge \tilde\w_{AdS}$. Therefore, for dual giants the bound becomes
\be
\int \left({\mathcal H} + p_{||} - \iota_0 C- \iota_{||} C \right)\geq \int - e^0 \wedge \tilde\w_{AdS}\label{calAdS}
\ee
We now show that the known dual giant configuration of Ref.~\cite{Grisaru:2000zn} saturates this bound.\\ 

We begin by writing the metric on $AdS_5$ slightly more explicitly. Recall from Eq.~(\ref{AdSmetric}) that this metric is given by
\[
ds^2 = -\cosh^2\rho\ dt^2 + d\rho^2 + \sinh^2\rho\ \sum_{i=1}^4 d\Omega_i^2 
\]
where $\sum_i \Omega_i^2 = 1$. We choose the following polar coordinates for $\Omega_i$, 
\begin{eqnarray*}
\Omega_1 = \cos\al_1 &\qquad& \Omega_2 = \sin\al_1\cos\al_2\\
\Omega_3 = \sin\al_1\sin\al_2\cos\al_3 &\qquad& \Omega_4 = \sin\al_1\sin\al_2\sin\al_3
\end{eqnarray*}
and we write $r = \sinh\rho$. In these coordinates, the $AdS_5$ metric becomes
\be
ds^2 = -(1+r^2)\ dt^2 + \frac{dr^2}{1+r^2} + r^2 (d\al_1^2  + \sin^2\al_1 d\al_2^2 + \sin^2\al_1\sin^2\al_2 d\al_3^2)
\ee
and the preferred time-like direction, $e^0$, defined in Eq.~(\ref{e0exp}), becomes
\be
e^0 = (1+r^2)\ dt - r^2 (\cos\al_2d\al_1 - \cos\al_1\sin\al_1\sin\al_2 d\al_2 + \sin^2\al_1\sin^2\al_2 d\al_3) \label{e0second}
\ee
From the probe calculations in Ref.~\cite{Grisaru:2000zn} it is known that there is a dual giant graviton which wraps a 3-sphere parameterised by $\al_1, \al_2, \al_3$ at fixed $r$. We denote the coordinates on the world-volume of this brane by $\sigma^\mu$ ($\mu=0, 1, 2, 3$) and here $\sigma^0= t$ (i.e. we choose static gauge) and $\sigma^ i = \al_i$. The dual giant graviton also moves on the surface of the $S^5$ along any equator. For concreteness, we take the motion on the sphere to be in the direction $\phi_1$ with $\mu_i$ fixed to the values $\mu_1=1$, $\mu_2, \mu_3 = 0$. Recall that the metric on the sphere is given by
\[
ds^2 = \sum_{i=1}^3 (d\mu_i^2 + \mu_i^2 d\phi_i^2)
\]
with the condition that $\sum_i \mu_i^2 = 1$.\\

We now calculate the quantities on the left hand side of the calibration bound Eq.~(\ref{calAdS}). To do this we must first calculate the Lagrangian for the dual giant graviton. As before, this is given by
\[
\lal = - \sqrt{-g} + {\mathcal P}(C^{(4)})
\]
The induced metric on the dual giant world-volume is
\be
ds^2 = (- 1-r^2 + \dot\phi_1^2)\ dt^2 + r^2 (d\al_1^2  + \sin^2\al_1 d\al_2^2 + \sin^2\al_1\sin^2\al_2 d\al_3^2)
\ee
where we have pulled back the $AdS_5\times S^5$ metric to the dual giant world-volume specified above. Therefore,
\[
\sqrt{ - g} = \sqrt {1+r^2 - \dot\phi_1^2}\ r^3 \sin^2\al_1 \sin\al_2
\]
The pull-back of the 4-form potential is
\be
{\mathcal P}(C^4) = C_{t\al_1\al_2\al_3} + \dot\phi_1 C_{\phi_1 \al_1\al_2\al_3} 
\ee
Hence, we obtain the following Lagrangian for the dual giant,
\be
\lal =  - \sqrt {1+r^2 - \dot\phi_1^2}\ r^3 \sin^2\al_1 \sin\al_2 + C_{t\al_1\al_2\al_3} + \dot\phi_1 C_{\phi_1 \al_1\al_2\al_3} 
\ee 
We can use this to calculate the momentum conjugate to $\phi_1$. We obtain,
\[
p_{\phi_1} = \frac{\partial \lal}{\partial \dot\phi_1} = \frac{r^3 \sin^2\al_1 \sin\al_2\ \dot\phi_1}{\sqrt {1+r^2 - \dot\phi_1^2}} +  C_{\phi_1 \al_1\al_2\al_3}
\]
Therefore, the Hamiltonian is
\be
{\mathcal H} = p_{\phi_1} \dot\phi_1 - \lal = \frac{r^3 \sin^2\al_1 \sin\al_2\ (1+r^2) }{\sqrt{1+r^2 - \dot\phi_1^2}} -  C_{t\al_1\al_2\al_3}
\ee
Recall that $e^{||} = \sum_i \mu_i^2 d\phi_i$, which on the dual giant world-volume reduces to $e^{||} = d\phi_1$. Therefore, $p_{||} = p_{\phi_1}$ and hence
\be
{\mathcal H} + p_{||} = r^3 \sin^2\al_1 \sin\al_2 \frac{1+r^2  + \dot\phi_1}{\sqrt{1+r^2 - \dot\phi_1^2}} - C_{t\al_1\al_2\al_3} +  C_{\phi_1 \al_1\al_2\al_3}
\ee
We now need to calculate $\int \left(\iota_0 C + \iota_{||} C\right)$. From the form of $e^0$ and the fact that $e^{||} = d\phi_1 $ on the trajectory, we obtain
\[
\int \left(\iota_0 C + \iota_{||} C\right) = \int \left(- C_{t\al_1\al_2\al_3} + C_{\phi_1\al_1\al_2\al_3}\right) d^3\alpha
\]  
Hence, 
\be
\int \left({\mathcal H} + p_{||} - \iota_0 C - \iota_{||} C\right) = \int r^3 \sin^2\al_1 \sin\al_2 \frac{1+r^2 + \dot\phi_1}{\sqrt{1+r^2 - \dot\phi_1^2}}\ d^3 \al \label{left}
\ee
which gives the left hand side of the bound.\\

The right hand side of the bound Eq.~(\ref{calAdS}) is given by
\be
\int - e^0 \wedge \tilde\w_{AdS}
\ee
We can calculate $\tilde\w_{AdS}$ easily since $d e^0 = -2 \tilde\w_{AdS}$. Using the expression for $e^0$ given in Eq.~(\ref{e0second}) we obtain,
\begin{eqnarray*}
\tilde\w_{AdS} &=& - r dr \wedge dt + r^2 \sin\al_1\cos\al_1\sin^2\al_2\ d\al_1\wedge d\al_3 \\
 &+& r^2 \sin^2\al_1\sin\al_2 (d\al_1\wedge d\al_2  + \cos\al_2\ d\al_2\wedge d\al_3)\\
&+& rdr \wedge (\cos\al_2 d\al_1 - \sin\al_1\cos\al_1\sin\al_2 d\al_2 + \sin^2\al_1\sin^2\al_2 d\al_3)
\end{eqnarray*}
Since the spatial world-volume of the dual giant is parameterised by $\al_1, \al_2, \al_3$, the right hand side of the bound is given by,
\be
\int - e^0 \wedge \tilde\w_{AdS} = \int  r^4 \sin^2\al_1 \sin\al_2\ d\al_1 \wedge d\al_2 \wedge d\al_3 \label{right}
\ee
Hence, from Eqs.~(\ref{left}) and (\ref{right}) the left and right hand sides of the bound Eq.~(\ref{calAdS}) are equal when $\dot\phi_1 = -1$. In the case where $\dot\phi_1 \neq -1$,
\[
\frac{1+r^2 + \dot\phi_1}{\sqrt{1+r^2 - \dot\phi_1^2}} > r 
\]
which means that the brane is not calibrated, but the bound Eq.~(\ref{calAdS}) is satisfied. In fact, for this brane the calibration bound is saturated if and only if $\dot\phi_1 =- 1, r=0$. Note that the speed $\dot\phi_1 =- 1$ agrees with the speed one obtains from a probe calculation \cite{Grisaru:2000zn}. Like giant gravitons, the centre of mass of the dual giant moves along a null trajectory. This can be seen by evaluating the $AdS_5\times S^5$ metric on the trajectory $\mu_1 =1, \dot\phi_1=- 1$ with $r = 0$, which corresponds to the centre of mass. Moreover, like the giant graviton, the surface elements of the brane move at less than the speed of light. The time-like trajectory taken by a surface element is simply $ds^2 = - r^2 dt^2$.\\


\section{Conclusions}
In this paper we have constructed $p$-forms from Killing spinors of type IIB supergravity. We find that non-zero 1-,3-,5-,7- and 9-forms can be constructed. Using the gravitino Killing spinor equation we have derived the full set of differential equations that the forms satisfy in a general supersymmetric background. In analogy to the 11-dimensional case, one combination of the 1-forms, namely $K^{11}+K^{22}$, is Killing. 
We have also derived some algebraic identities satisfied by the forms using the algebraic Killing spinor equation and Fierz identities. It is interesting that the Fierz identities force $K\cdot K=\Phi\cdot \Phi =\Sigma\cdot \Sigma =0$. This is different to the case in 11 dimensions, where the Killing vector, $K$, can be time-like or null. However, here in 10 dimensions only the null case is allowed. The differential and algebraic relations we have derived could now be used for classifying general supersymmetric type IIB backgrounds using the ideas of G-structures. However, one complication in 10 dimensions is that there are four independent background field strengths, so classifying the most general backgrounds might be more difficult than the 11-dimensional case.\\

The second aspect of this paper has been to consider non-static branes from the point of view of calibrations. This is an interesting problem to consider because most previous work on calibrations has focused on static probe branes, even though non-static branes are known to play an important role in supergravity/string theory (e.g. giant gravitons are important in the AdS/CFT correspondence). We have given a concrete example of a non-static brane, namely a giant graviton in $AdS_5\times S^5$. We have found the calibration bound that these branes saturate, minimising $\int K\cdot p\ $. The minimised quantity corresponds to ``energy minus momentum'' in this case. Moreover, dual giant gravitons also saturate the calibration bound and minimise the same quantity. It would now be interesting to consider the 2-spin giant gravitons introduced in Ref.~\cite{Arapoglu:2003ti} from the point of view of calibrations. There are also many other examples of non-static branes which one could consider, e.g. supertubes. It would be interesting to understand these branes using calibrations.

\vspace{1cm} \noindent {\bf\Large Acknowledgements}\\

We would like to thank David Page, Hannu Rajaniemi and Simon Ross for helpful discussions. EJH is supported by the University of Adelaide and the Overseas Research Students Awards Scheme.

\end{document}